\documentstyle[prl,twocolumn,epsf,epsfig,aps]{revtex}
\newcommand{\BEQ}{\begin{equation}}
\newcommand{\EEQ}{\end{equation}}
\newcommand{\BEA}{\begin{eqnarray}}
\newcommand{\EEA}{\end{eqnarray}}

\renewcommand{\d}{{\rm d }}

\newcommand{\I}{{\cal I}}
\newcommand{\p}{\partial}

\newcommand{\halv}{\mbox{$\frac{1}{2}$}}
\newcommand{\kvart}{\mbox{$\frac{1}{4}$}}

\input{psfig}
\draft

\begin{document}



\title{Competition between glassiness and order in a multi-spin glass}

\author{J. A. Hertz$^a$, David Sherrington$^b$ and Th. M. Nieuwenhuizen$^c$}


\address{
a: NORDITA, Blegdamsvej 17 , DK-2100 Copenhagen, Denmark
\\ 
\noindent b: Department of Physics, University of Oxford,
Theoretical Physics, 1 Keble Road, Oxford, OX1 3NP, England
\\ 
\noindent c: Department of Physics and Astronomy,
 Valckenierstraat 65, 1018 XE Amsterdam,
  The Netherlands. }
\date{Version 29 April 1999; printout  \today}
\maketitle
\begin{abstract}
A mean-field multi-spin interaction spin glass model is analyzed in the
presence of a ferromagnetic coupling.  The static and dynamical phase
diagrams contain four phases (paramagnet, spin glass, ordinary ferromagnet 
and glassy ferromagnet) and exhibit reentrant behavior.  The glassy
ferromagnet phase has anomalous dynamical properties.  The results are
consistent with a nonequilibrium thermodynamics that has been proposed
for glasses. 
\end{abstract}
\pacs{04.70-s, 04.70 Dy, 05.70-a, 71.55 Jv, 97.60 Lf}



Recent work has emphasized the importance of aging as a fundamental
property of glassy systems and given it a precise meaning:  While simple
average quantities relax relatively quickly to stationary values,
two-time quantities (correlation and response functions) show that the
system never truly equilibrates.   Whenever the separation between the
two times is comparable with the age of the system or greater, time
translational invariance is violated and the fluctuation-dissipation
relation (FDR) is modified.  These effects are a direct consequence of
trapping in metastable attractors, and the modified FDR gives 
information about the overlap distribution of these attractors.  This 
has been demonstrated explicitly for several soluble models 
\cite{CK,FM,CKLD-CLD}. 

However, in all these models the distribution of the quenched random 
variables is symmetric, so the metastable attractors lack interesting
macroscopic structure.  This is not the case in a large variety of systems
where biased, tuned, or trained interactions lead to cooperatively ordered
attractors, yet this macroscopic ordering competes with significant quenched 
randomness and the consequent tendency toward glassiness.  Examples of such 
systems include models for recurrent neural networks \cite{AGS}, error 
correction algorithms \cite{Sourlas}, combinatorial optimization 
\cite{FA,BSS,Remiksat}, and proteins \cite{protein}, as well as 
experimental spin glass materials with ferromagnetic, antiferromagnetic or
helically ordered phases \cite{Mydosh}.  Equilibrium analyses of many of 
these have found replica symmetry breaking regions, indicative of glassy 
behaviour, in their phase diagrams, but so far their dynamics have not 
been studied. 

The purpose of this paper is to start to remedy this lack by solving
a nontrivial model with both spin-glass-like and macroscopic attractors. 
We demonstrate that (i) the ferromagnetic part of the dynamical phase diagram 
contains both glassy and ordinary non-glassy regions (with aging present in
 the glassy one), (ii) the two-time correlation function in the glassy 
ferromagnet involves non-analytic features absent in the zero-field spin-glass 
\cite{CKLD-CLD}.  The results are in accord with a proposed non-equilibrium 
thermodynamic description of glasses \cite{Nthermo,NEhren,Nhammer}.

For the model we discuss (a spherical p-spin glass \cite{CK,CHS,CS,Saak} 
with a ferromagnetic interaction), we observe several interesting 
features of the phase diagrams (Fig.\ 1). (i) For both statics and dynamics, 
they contain four phases: paramagnet, spin glass (with zero spontaneous
magnetization), conventional ferromagnet (non-zero spontaneous magnetization
but no replica symmetry breaking or aging), and glassy ferromagnet (non-zero 
spontaneous magnetization with replica symmetry breaking for statics and
aging for dynamics).  All the phases meet at a multicritical point. (ii) The 
critical ferromagnetic exchange separating the spin-glass and glassy 
ferromagnetic regions decreases with increasing temperatures, so that within 
a finite band of exchange interaction values there occurs a sequence
 of phases, 
with decreasing temperature, of paramagnet, ordinary ferromagnet, glassy 
ferromagnet and spin glass.  This has been a regularly observed feature of 
experiment (referred to as ``re-entrance'' \cite{Mydosh}), but is not 
found in equilibrium 
theory for conventional spin-glass models \cite{SKcomm}. (iii) There is a 
finite maximum ferromagnetic exchange for glassy ferromagnetism even at 
zero temperature for all finite $p>2$\cite{Comment1}. 
(iv) The transition temperature 
separating glassy and ordinary ferromagnetic  regions first rises and then 
falls as the ferromagnetic exchange is increased beyond its value at the 
multicritical point, thereby indicating re-entrance as the ferromagnetic 
exchange is increased within an appropriate temperature band, with the 
sequence paramagnet, ordinary ferromagnet, glassy ferromagnet, and back to 
ordinary ferromagnet.  (v) The peak of the phase line separating glassy and 
ordinary ferromagnet also marks a boundary between two types of onset of 
one-step replica symmetry breaking (1RSB), discontinuous for smaller 
ferromagnetic exchange, continuous for larger ferromagnetic exchange;
 $cf$.\cite{CHS}.  
(vi) Wherever the onset of 1RSB is discontinuous, the dynamical transition 
temperature is higher than the static one, as in the limit of zero 
ferromagnetic interaction \cite{CHS,CS,Marginality}.
  
We consider a model Hamiltonian
\begin{eqnarray}
{\cal{H}}&=&-\sum_{i_{1}<i_{2}..<i_p} J_{i_{1}i_{2}..i_{p}}
S_{i_{1}}S_{i_{2}}..S_{i_{p}}					\nonumber \\
& -& \frac{J_0}{N}\sum_{ij}
S_{i}S_{j} -    H\sum_i S_i  			\label{eq:Hamiltonian}
\end{eqnarray}
with independently distributed random quenched $p$-spin interactions of mean 
zero and variance $J^2p!/2N^{p-1}$ and nonrandom 2-spin interactions.  The 
spins are subject to the spherical constraint $\sum_iS^2_i=N$.  Mean field 
theory is exact for infinite-ranged interactions.    The choice of 
spherical spins simplifies the resulting self-consistency equations, while 
$p>2$ ensures that one-step replica-symmetry breaking (1RSB) is sufficient. 

\begin{figure}[bhb]
\vspace{-0.5cm}\hspace{-1.5cm}
\vbox{\hfil\epsfig{figure=
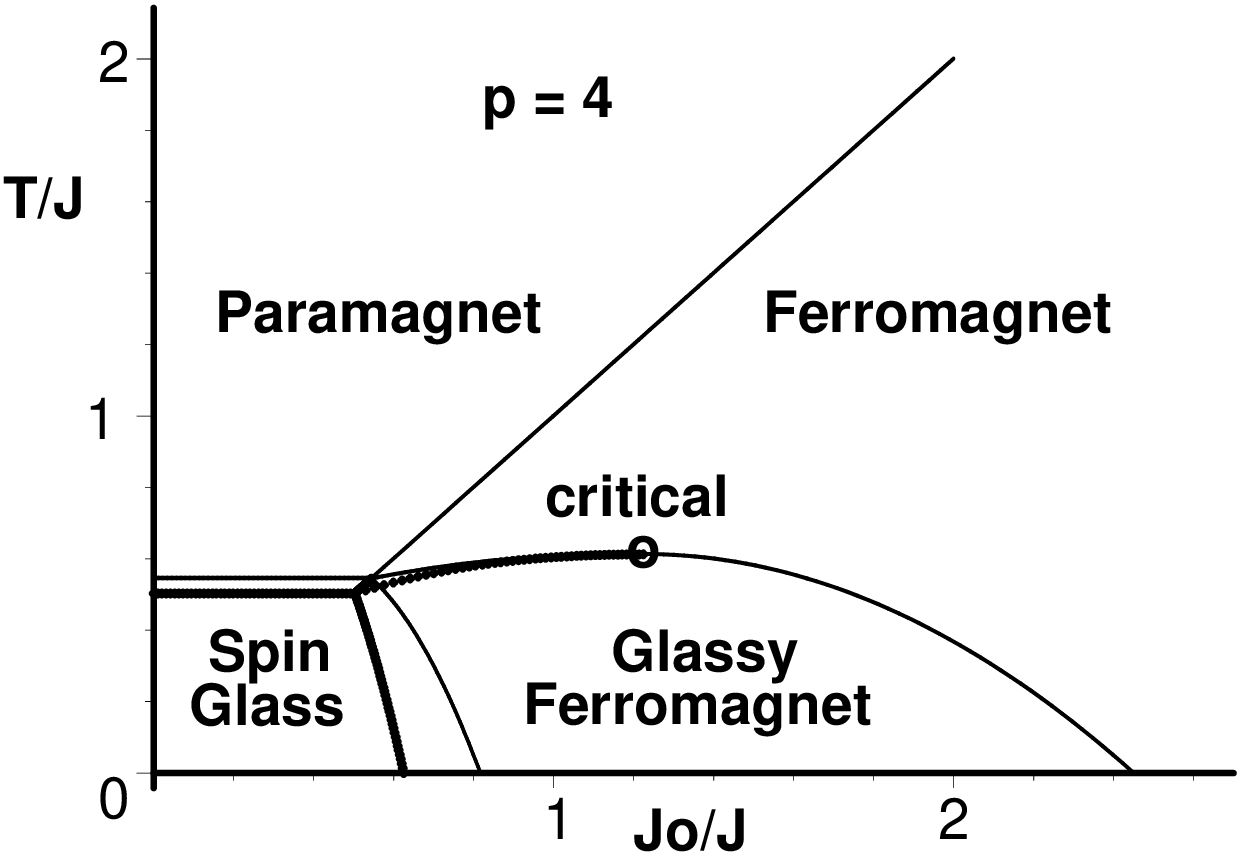,width=8cm,angle=270}\hfil}  
\vspace{-1.75cm}
\caption{Static and dynamic phase diagram for the model with $p=4$.
When different from the dynamical ones, 
the static phase boundaries are indicated by bold lines.}
\end{figure}
 
We have studied the model by two complementary approaches.  The first employs 
the replica formalism and permits us to obtain both the equilibrium 
and dynamical order parameters.  
It is characterized by three order parameters the maximum 
(self-) overlap $q_1$, the minimum (mutual) overlap $q_0$, the magnetization 
$M$ and the amplitude $(1-x)$ of the self-overlap part of the overlap 
probability distribution.  The spherical constraint is ensured by a
self-consistently determined Lagrange multiplier.  Stationarity of the
replica free energy 
\BEA 
F &=& -\halv J_0M^2 
- \kvart \beta J^2 [1-(1-x) q_1^p-xq_0^p] \nonumber 	\\
&-& HM  -\halv (T/x)\log [1-(1-x) q_1-xq_0]	 \nonumber  	\\
&+&   \frac{(1-x)T}{2x}\log(1-q_1)				  
+   \frac{(M^2-q_0)T}{2(1-(1-x) q_1-xq_0)}\label{eq:bFCS}
\EEA
with respect to $q_0$, $q_1$ and $M$ yields the 
self-consistency equations
\BEA
M &=& (\beta H+\beta J_0 M)(1-\overline{q}) 		\label{eq:Meqn} \\
q_0 &=& \mu(1-\overline{q})^2q_0^{p-1}+M^2		\label{eq:$q_0$eq} \\
q_1 - q_0 &=& \mu(1-\overline{q})(1-q_1)(q_1^{p-1} -
q_0^{p-1}),						\label{eq:$q_1$eq}
\EEA
where we have used the shorthands $\mu={1\over 2} p\beta^2J^2$ and 
$\overline{q}=xq_0+(1-x)q_1$.

For the equilibrium (static) theory, a fourth self-consistency condition
is provided by requiring that the derivative
 \begin{eqnarray}
\frac{\partial F}{\partial x} = & & \frac{T}{2}\left[
\frac{1}{x^2}\log\frac{1-\overline {q}}{1-q_1}
-\frac{q_1-q_0}{x(1-\overline{q})} \right.		\nonumber \\ 
&-& \left. \frac{\beta^2J^2}{2}(q_1^p-q_0^p) 
-\frac{(M^2-q_0)(q_1-q_0)}{(1-\overline{q})^2}\right]	\label{eq:xstatic}
\end{eqnarray}
vanish. 
Eqns.\ (\ref{eq:Meqn}-\ref{eq:xstatic}) are then solved for $q_0$, $q_1$,
$M$ and $x$.

To obtain the dynamical order parameters one employs, instead of Eqn.\ 
(\ref{eq:xstatic}), the marginal stability condition 
\cite{CHS,Marginality,Nmaxmin}
\begin{equation}
(p-1)\mu q_1^{p-2}(1-q_1)^2=1 .				\label{eq:xdynamic}
\end{equation}
As in  the problem without a ferromagnetic term \cite{CHS}, this procedure 
yields the same order parameters and transitions that we find with our 
second approach, a direct dynamical analysis. 
 
That treatment starts from the Langevin equation
\begin{equation}
\frac{\partial S_i}{\partial t} = -\frac{\partial{\cal{H}}}{\partial S_i} 
-z(t) S_i +\eta_i(t)					\label{eq:Langevin}
\end{equation}
where $\eta_i(t)$ is white noise of temperature $T$ and
$z(t)$ has to be adjusted to satisfy the spherical constraint.
Following and extending now-standard procedures \cite{SZ} of introducing
a generating functional, averaging over stochastic noise and
quenched disorder, introducing appropriate macroscopic time-dependent
quantities and using extremal analysis in the limit $N\to \infty$,
there result self-consistent equations for the local correlation
function $C(t,t^\prime)=(1/N)\sum_i \langle S_i(t)S_i(t^\prime)\rangle$,
the local response function $G(t,t^\prime)=(1/N)$
$\sum_i \delta \langle S_i(t)\rangle/\delta H_i(t^\prime)|_{H_i(t^\prime)=H}$,
and the global magnetization $M(t)=(1/N)\sum_i\langle S_i(t)\rangle$:
\begin{eqnarray}
&\partial_t &C(t,t^\prime)=-z(t)C(t,t^\prime) + 2 G(t^\prime,t)
+\beta H M(t')						\nonumber	\\
&+&\beta J_0 M(t)M(t^\prime)+
\mu \int_0^{t'} \d t_1 \,C^{p-1}(t,t_1)G(t',t_1)	 \nonumber 	\\
&+&(p-1)\mu \int^t_0 dt_1\,G(t,t_1)C^{p-2}(t,t_1)C(t_1,t^\prime) 
 \label{eq:Csc}
\end{eqnarray}
\begin{eqnarray}
&\partial_t&G(t,t^\prime)= -z(t)G(t,t^\prime)+\delta(t-t^\prime) \nonumber \\
&+&(p-1)\mu\int^t_{t^{\prime}}\d t_1 \,G(t,t_1)C^{p-2}(t,t_1)G(t_1,t^\prime)
								\label{eq:Gsc} 
\end{eqnarray}
\begin{eqnarray}
&\partial_t&M(t)=-z(t)M(t)+\beta H+ \beta J_0 M(t) 		\nonumber  \\
&+&(p-1)\mu\int^t_0 \d t_1\,
G(t,t_1)C^{p-2}(t,t_1) M(t_1)					\label{eq:Msc}
\end{eqnarray}
Together with the spherical constraint $C(t,t) =1$, Eqs.\ 
(\ref{eq:Csc}-\ref{eq:Msc}) determine the dynamics completely.  However, 
even for $J_0=H=0$, they have not yet been solved.  We concentrate on  
long times, where $z(t)$ and $M(t)$ reach stationary values and a 
self-consistent solution is possible under the assumption that 
$C(t,t^\prime)$ and $G(t,t^\prime)$ have time-translation-invariant 
behavior for $(t-t^\prime) \ll t^\prime$ and simple aging behavior for 
$(t-t^\prime) \gg t^\prime$ \cite{CK}:
\BEA
C(t,t^\prime)&=&C_{st}(t-t^\prime) + C_{ag}(t,t^\prime), 	 \\
G(t,t^\prime)&=&G_{st}(t-t^\prime)+G_{ag}(t,t^\prime), 		 
\EEA
where in terms of $\lambda=t^\prime/t$, $C_{ag}(t,t^\prime)={\cal{C}}(\lambda)$
and $G_{ag}(t,t') = {\cal G}(\lambda)/t$, with limiting values 
$C_{st}(0)=1-q_1$, $C_{st}(\infty)=0$, ${\cal{C}}(1)=q_1$, and 
${\cal{C}}(0)=q_0$.  Thus $q_1$ is the plateau value of $C$ reached for 
$1 \ll t-t' \ll t'$, and $q_0$ is its asymptotic ($t \rightarrow \infty$) 
limit.  In the stationary regime the conventional 
fluctuation-dissipation theorem
\begin{equation}
\frac{\p C_{st}(t-t')}{\p t'} =T\,G_{st}(t-t') \qquad(t>t') 	\label{eq:FDT} 
\end{equation}
holds, while in the aging regime one has instead
 the modified fluctuation-dissipation relation
\begin{equation}  
\frac{\p C_{ag}(t,t')}{\p t'}=T_e\,G_{ag}(t,t')	 
\qquad T_e=\frac{T}{x},					\label{eq:FDT2}
\end{equation}
i.e.\ $\d{\cal C }(\lambda)/\d\lambda =T_e{\cal G}(\lambda)$.
 
At long times the time-derivative terms on the left-hand sides of
(\ref{eq:Csc}-\ref{eq:Msc}) can be neglected.  The asymptotic solution 
found in this limit for the aging regime admits a reparametrization 
invariance \cite{Sompolinsky}: if $C(t,t')$ is a solution, so is 
$C(h(t), h(t'))$, with $h(t)$ an arbitrary monotonic function of $t$.  
Thus we cannot find the complete time-dependence of $C$ and $G$ from 
the asymptotic equations alone.  Nevertheless, we can solve for $M$, $x$, 
$q_0$ and $q_1$, finding the same results as were obtained above by the 
replica treatment in its `dynamics' form.  No further assumptions on 
$C$ and $G$ are needed to obtain these results.  
 
We restrict ourselves to $H=0$ in this letter.
Fig.\ 1 shows the phase diagram of the model in $T-J_0$ space for $p=4$.
The general features are not sensitive to $p$.  For sufficiently weak
ferromagnetic interaction $J_0$, we find the same results as for $J_0=0$:
a dynamical paramagnetic-to-spin glass transition at a temperature $T_d$
and a static transition at a lower temperature $T_g$.  The spin glass
states (both dynamical and equilibrium) involve 1RSB,
with $q_0 = M = 0$, and $q_1$ is discontinuous at the transitions, where
$x \rightarrow 1$.  For $J>T_d$ ($T_g$ for statics), there is a Curie
temperature $T_c$, below which the paramagnetic state is unstable against
the onset of spontaneous magnetization. For a range of temperatures
below $T_c$ this ferromagnetic state has
a nonzero spin glass order parameter, but no glassy properties (no RSB
or aging): $q_1 = q_0$.  However, it is unstable, at low enough
 temperatures and
not too large $J_0$, against the formation of a glassy ferromagnetic
state with nonzero $M$, $q_0$, and $q_1 > q_0$ (which implies aging).  Below a
temperature-dependent critical value of $J_0$, the ferromagnetism of this
state becomes unstable, and one recovers the simple spin glass phase.
All four phases come together at the point $J_0 = T = T_d$ ($J_0=T=T_g$
for statics).

The upper boundary of the glassy ferromagnetic phase rises as $J_0$
increases from $T_d$ ($T_g$ for statics) and reaches a maximum at
$J_0/J = \sqrt{p(p-1)/2} [(p-2)/p]^{(p-2)/2}$.  It falls to $T=0$
at $J_0/J = \sqrt{p(p-1)/2}$.   To the left of the maximum, $q_1-q_0$
jumps discontinuously and $x \rightarrow 1$ at the transition (as in
the paramagnetic-to-spin glass transition at small $J_0$).  To the
right of the maximum, $q_1-q_0$ goes continuously to zero at the
transition, where $x<1$, and the static and dynamical boundaries coincide.
This part of the boundary is thus an Almeida-Thouless line like that found
in the SK model \cite{AT}, though the low-$T$ states are different:
here one step of RSB is exact, while full RSB is necessary in the SK
model.  The overall shape of the boundary is similar to that in the 
p-spin glass in an external field \cite{CHS,CS}, though 
the relation between the two phase diagrams is not elementary because
the order parameter equations are coupled.  

In the large-$p$ limit, the dynamical small-$J_0$ spin glass transition
temperature $T_d$ approaches a finite limit $1/\sqrt{2{\rm e}}$.  The
glassy ferromagnetic phase extends out to $J_0 = Jp/\sqrt{2}$, with its
maximum temperature, $2T_d/\sqrt{\rm e}$ achieved at
$J_0 = Jp/({\rm e}\sqrt{2})$.  The {\em static} spin glass transition
temperature $T_g$ at small $J_0$ goes to zero like $1/\sqrt{\log p}$
\cite{Saak}, so
the spin glass phase disappears, but the glassy ferromagnetic
phase remains.
 
\begin{figure}
\vspace{-0.2cm}\hspace{-2cm}
\vbox{\hfil\epsfig{figure=
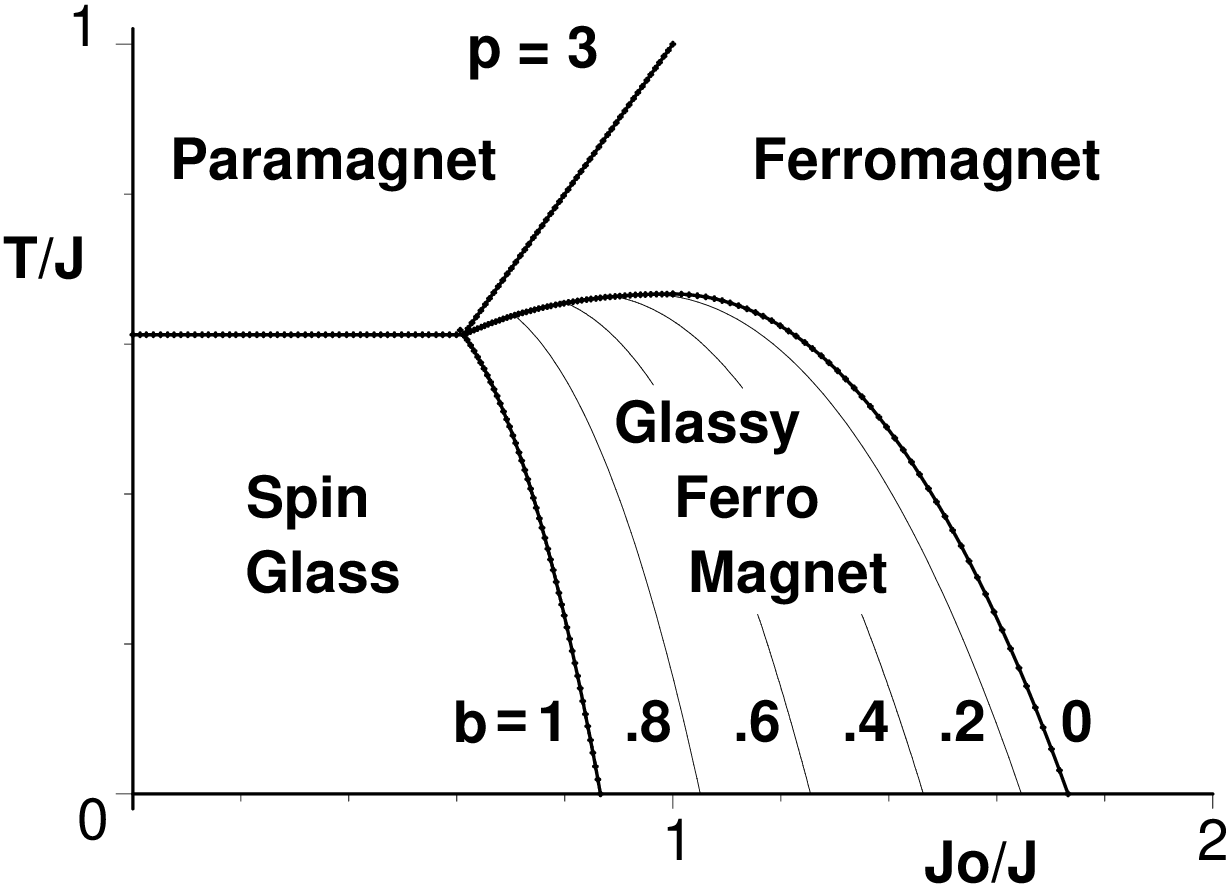 ,width=8cm,angle=270}\hfil} 
\vspace{-2cm}
\caption{
Dynamical phase transition lines (solid)
for the model with $p$$=$$3$, and 
lines of constant $b$ in the glassy ferromagnet.  }
\end{figure}
The glassy state is characterized by a
plateau $C=q_1$ in the correlation function, the length of which is
age-dependent.
The behavior near this plateau involves 
power laws.  An exponent $a$ characterizes the decay 
at the end of the short-time regime (the approach to the plateau):
$C(t+\tau,t)\equiv C(\tau) \approx q_1 + {\rm const}/\tau^{a}$.  Using Eqn.\ 
(\ref{eq:Gsc}) in the FDT regime, we find an equation for $a$,
\begin{equation}
\frac{\Gamma^2(1 - a)}{\Gamma(1 - 2a)} 
= \frac{(p-2)(1 - q_1)}{2q_1}, 					\label{eq:a}	
\end{equation}
valid in both glassy ferromagnetic and spin glass phases.  Thus, 
$a$ is independent of $J_0$, since $q_1$ is
fixed by (\ref{eq:xdynamic}).

For $J_0=0$ the asymptotic dynamical equations have the exact aging
solution ${\cal C}(\lambda) = \lambda^{\nu}$ ($0 < \nu < 1$)
\cite{CK}, which also holds for $J_0 > 0$ in the spin glass phase. 
In the glassy ferromagnetic phase, ${\cal C}(\lambda)$ is nonanalytic
for $\lambda \rightarrow 1$, i.e., at the end of the plateau and the
beginning of the aging regime.  We make the Ansatz
${\cal C}(\lambda) = 1 - B(1-\lambda)^{b} + {\cal O} [(1-\lambda)^{2b}]$
and find, using the result (\ref{eq:a}), that the exponent $b$ must satisfy
\begin{equation}
x\frac{\Gamma^2(1+b)}{\Gamma(1+2b)} =
\frac{\Gamma^2(1-a)}{\Gamma(1-2a)},				\label{eq:b}
\end{equation}
Such an Ansatz was also employed in a different problem \cite{CKLD-CLD}, 
where the relation (\ref{eq:b}) was also found.  A
similar result is expected to hold for a spin glass in a field, as well.
At the boundary between the spin glass and glassy ferromagnetic
phases, $b \rightarrow 1$, and along the AT line separating the
conventional and glassy ferromagnetic phases $b \rightarrow 0$.
Fig.\ 2 shows lines of constant $b$ for the $p=3$ model.
 
The system dynamically condenses into glassy states,
characterized by $x$ or the effective temperature $T_e=T/x$.
They  have a configurational entropy (or complexity)
~\cite{Nthermo,NEhren,CS95,Monasson}
$\I = -(\p F /\p T_e)_T = (x^2/T) (\p F/\p x)_T$,
that follows from Eqn.\ (6).
$\I$ is 
a positive constant in the spin glass
\cite{CS95}, and, of course, it is zero in the ordinary ferromagnet.
In the glassy ferromagnet it interpolates smoothly between these
two values, and it vanishes at the transition to the ordinary ferromagnet, 
where $q_0\to q_1$.  

The specific heat, defined as the limit of
the energy difference between states obtained by rapid quenches to two 
slightly different temperatures, divided by the temperature difference, has 
two terms: $C=\d U/\d T= T \d S_{1}/\d T+
T_e \d \I/\d T$
\cite{Nthermo,NEhren}, with the intravalley entropy given by
$ S_{1}=-(\p F/\p T)_{T_e}$, viz.
\BEQ
 S_{1}= 
\halv \ln(1-q_1)-\kvart (\beta J)^2 (1+(p-1)q_1^p-pq_1^{p-1}).
\EEQ
This is exactly the entropy of a single TAP valley~\cite{CS95,KPV} and 
describes states that can be reached dynamically at fixed $T$. 
In the spin glass and in the ordinary ferromagnet
$\d \I/\d T=0$, so the specific heat merely follows from the 
intravalley processes.
In the glassy ferromagnet, however, $\d \I/\d T<0$, 
so the specific heat (as defined here) acquires a (negative)
contribution from changes in the shape of the free energy landscape with 
temperature.    

In summary, we have been able to elucidate explicitly the consequences of
the competition between glassiness and ferromagnetic ordering in the 
statistical mechanics and long-time dynamics of an asymptotically soluble 
model.  We have found several novel features and extended and
verified the applicability of concepts devised for spin glasses. 
These results can shed useful light on the many other important problems,
in physics and other fields, where ordering competes with quenched
disorder.

\acknowledgements

The authors are grateful to A. Cavagna, I. Giardina, and H. Horner
for for stimulating discussions.
Hospitality at the Universities of Oxford (J. A. H.) and
Leuven (Th. M. N.), and at the ICPT, Trieste,
is also acknowledged.


\begin{thebibliography}{ppppp}
\bibitem{SKcomm} Re-entrance was found in the original
replica-symmetric but unstable solution of the SK model \cite{SK}.
\bibitem{Comment1} This is to be contrasted with the situation for 
SK or $p=\infty$ models  \cite{Derrida},
where glassy ferromagnetism  extends to infinite $J_0$.
\end{thebibliography}
\end{document}